\documentclass[onecolumn]{emulateapj}

\usepackage{graphicx}
\usepackage{natbib}
\usepackage{amsmath, amsthm, amssymb}

\begin{document}

\title{Reconstructing the shape of the correlation function}
\author{K.~M.~Huffenberger, M.~Galeazzi, E.~Ursino}

\affiliation{Department of Physics, University of Miami, Coral Gables, Florida 33146}

\begin{abstract}
We develop an estimator for the correlation function which, in the ensemble average, returns the shape of the correlation function, even for signals that have significant correlations on the scale of the survey region.  Our estimator is general and works in any number of dimensions.  We develop versions of the estimator for both diffuse and discrete signals.  As an application, we examine Monte Carlo simulations of X-ray background measurements.  These include a realistic, spatially-inhomogeneous population of spurious detector  events.   We discuss applying the estimator to the averaging of correlation functions evaluated on several small fields, and to other cosmological applications.
\end{abstract}


\section{Introduction}

Two-point statistics encode valuable information about the fields that they describe, such as the cosmological matter density traced by galaxies or the intensity of radiation in backgrounds like the Cosmic Microwave Background (CMB), the Cosmic Infrared Background (CIB), or the Diffuse X-ray Background (DXB).

For discrete objects, the two-point, dimensionless correlation function can be defined in terms of the probability of finding a pair of objects in two small cells, with sizes $\delta\Omega_1$ and $\delta\Omega_2$, separated by $\theta_{12}$ \citep[][\S 31, 45]{1980lssu.book.....P}:
\begin{equation}
  \delta P_{12} = {\cal N}^2 \delta \Omega_1 \delta\Omega_2 \left[1+w(\theta_{12}) \right]
\end{equation}
where ${\cal N}$ is the mean density of sources.
For diffuse fields, the equivalent definition for a signal $s$ with mean $\langle s \rangle = \mu$ is
\begin{equation}
  \langle s_1 s_2 \rangle = \mu^2 \left[ 1 + w(\theta_{12}) \right],
\end{equation}
where here and throughout $\langle \dots \rangle$ denotes the ensemble average.\footnote{If the signal $s$ records the object count in a cell with size $\delta \Omega$, then $\langle s \rangle = \mu={\cal N}\delta\Omega$.  If the cells are so small that they contain at most one object,  $\langle s_1 s_2 \rangle = \delta P_{12}$, making the correspondence between the two definitions clear.}  We denote the covariance of $s$ as $C(\theta) = \mu^2 w(\theta)$, which we also refer to as the (dimensionful) correlation function.  This work mostly deals with the dimensionful correlation function and addresses the bias in its estimation.  With similar expressions, we can define correlation functions in any number of dimensions, replacing the angular separation $\theta$ by a linear separation or time interval or whatever is appropriate.


The estimation of the correlation function has been studied extensively in the literature.  For galaxy clustering, \citet{1982MNRAS.201..867H}, \citet{1983ApJ...267..465D}, and \citet{1993ApJ...417...19H} suggest different Monte Carlo estimators, but the most common estimator now in use was advocated by \citet{1993ApJ...412...64L}, which employs the data in concert with a synthetic, random catalog.  Their estimator combines counts of objects pairs within and between the data and random catalogs.  
This estimator is biased, but for surveys where the correlation length of the objects is much smaller than the survey area, the bias is small  \citep{1994ApJ...424..569B}.  Such is the case for modern galaxy surveys like 2dF and SDSS \citep{2001MNRAS.327.1297P,2000AJ....120.1579Y}.  However, the bias can become significant when structures approach the size of the survey \citep{1999A&A...343..333K}.  This bias can be corrected \citep[e.g.][]{2002ApJ...579...48S}, but the correction depends on same correlation function that is being estimated.

For diffuse signals like the CMB, where using the dimensionful correlation function is more common, a typical estimator looks like \citep{1996ApJ...464L..25H,2007PhRvD..75b3507C}:
\begin{equation}
\tilde C_0(\theta) =  \frac{\sum_{ij}\alpha_i\alpha_j(s_i-\tilde \mu)(s_j - \tilde \mu)}{ \sum_{ij} \alpha_i\alpha_j} \label{eq:intro_continuous_C0}
\end{equation}
where $\alpha_i$ are the weights applied to the pixels or cells (for the purpose of downweighting noisy regions), $\tilde \mu$ is an estimate of the mean, and the sum over $ij$ refers to pixels separated by $\theta$.  These estimators suffer the same biases on small fields.


In this paper we introduce a new method to address the biases in these above estimators.  Our estimator is also biased, but biased in a particularly convenient way: regardless of the survey geometry or weighting, the shape of the correlation function is preserved on average, and only information about a constant offset is lost.  This permits the straightforward averaging of correlation functions from several small patches across the sky.  Building upon the estimator in eqn.~(\ref{eq:intro_continuous_C0}), we develop classes of estimators for both diffuse signals and discrete objects. 

This work was prompted by our group's efforts to compute correlation function from observations of the diffuse X-ray background.
  The signal in that case comes from a diffuse, gaseous source, but arrives and is recorded as individual, discrete X-ray photons, and so can be analyzed with either scheme above.  Indeed, for simulations of diffuse X-ray emission from the WHIM, \citet{2011MNRAS.414.2970U} found that the \citet{1993ApJ...412...64L} estimator gave roughly equivalent results to an estimator of the type in eqn.~(\ref{eq:intro_continuous_C0}).  We focused on the correlation function biases because the angular correlation scale of this gas (several arcminutes) is substantial compared to the field-of-view ($\sim 8$ arcminutes) for single-field observations with the Chandra X-ray Observatory. 

The paper is organized as follows.  In section \ref{sec:continuous} we find the bias for the naive estimator (eqn.~\ref{eq:intro_continuous_C0}),  verifying our result with Monte Carlo simulations, and introduce a method for correcting it up to a constant offset.  In section \ref{sec:poisson} we extend this estimate to Poisson-distributed counts, allowing for the possibility of a spatially-varying set of spurious detector events.  Finally, we summarize our conclusions in section \ref{sec:conclusions}.  An appendix contains the detailed derivations of the bias terms.

\section{Correlation function estimator bias}\label{sec:continuous}

We begin by defining our signals.  
Let $s_i$ represent a pixelized, diffuse signal that is statistically homogeneous and isotropic.  Let it be described by a mean and covariance as follows:
\begin{eqnarray}
  \langle s_i \rangle &=& \mu  \\ \nonumber
  \langle (s_i-\mu)(s_j - \mu) \rangle &=&   \langle s_is_j  \rangle -\mu^2 = C(\theta_{ij} )
\end{eqnarray}
where 
$\theta_{ij}$ represents the separation between cells $i$ and $j$. In our derivations we use $C(\theta)$ rather that $w(\theta)$ because the examination of biases is convenient; $C(\theta)$ also makes sense for diffuse fields where $\mu = 0$.
No other special properties of $s$ are required, except that the covariance matrix is positive semi-definite: $ 0 \leq |C(\theta)| \leq C(0)$.  In particular, the signal need not be a Gaussian random field: we could define higher-order moments without disrupting our following arguments.  Note that by this definition, the correlation function $C(\theta)$ is a property of the probability distribution for our signal $s$, and it is not a descriptive statistic.


With a set of weights on the pixels, $\alpha_i$, we can compute a weighted average to estimate the mean,
\begin{equation}
  \tilde\mu = \frac{\sum_i \alpha_i s_i}{\sum_i \alpha_i}
\end{equation}
where the sum is over all pixels.  These weights could be chosen to be uniform or to suppress noisy or polluted portions of the measurement. Throughout we mark estimated quantities with tildes.
This mean estimate is unbiased, $\langle \tilde\mu \rangle = \mu$.  Additionally we define the deviation between the true mean and the estimated mean by
\begin{equation}
  \delta\tilde\mu = \tilde\mu - \mu
\end{equation}
with $\langle \delta\tilde\mu \rangle = 0$.

\subsection{Naive correlation function estimator}

Based on the estimated mean, we make an initial estimate of the correlation function in a bin labeled by $\theta_p$, which we call the naive estimator:
\begin{equation}
  \tilde C_0(\theta_p) = \frac{ \sum_{ij} d_{ij}(\theta_p) \alpha_i \alpha_j (s_i-\tilde\mu)(s_j - \tilde\mu)}{\sum_{ij} d_{ij}(\theta_p) \alpha_i \alpha_j}
  \label{eq:Cestimator}
\end{equation}
This is just a more explicit rewriting of eqn.~(\ref{eq:intro_continuous_C0}).
The function
\begin{equation}
  d_{ij}(\theta_p) = \left\{ 
    \begin{array}{ll}
      1, & \mbox{if $i$ and $j$ are separated by $\theta_p \pm \delta\theta$/2} \\
      0, & \mbox{otherwise}
    \end{array} \right.
\end{equation}
chooses the separation bin to which the pixel sum contributes.\footnote{$d_{ij}(\theta)$ is equivalent to the $\Theta_{ij}^\theta$ function defined by \citet{1993ApJ...412...64L}.  In practice we loop over all pixels and just select which separation bin is appropriate to accumulate the sum.} Evaluation of the estimator costs ${\cal O}(N^2)$ operations over $N$ pixels. If the true mean $\mu$ replaces the estimated mean $\tilde\mu$ in eqn.~(\ref{eq:Cestimator}), then this correlation function estimate is unbiased,\footnote{Technically, biases are also introduced by averaging the smooth sky into pixels---this pixel window function is severe if the pixels approach the size of the correlation length---and by binning the smooth correlation function into a stepwise function.  These can often be made insignificant by choosing finer discretization schemes, and we do not treat such biases here.}  and we find $\langle \tilde C_0(\theta_p) \rangle = C(\theta_p)$. However, since we do not know the true mean, our estimate will be biased, because we are forced to use the same (correlated) set of pixels to compute the mean and the correlation function.  The smaller the survey compared to the correlation length of the signal, the worse this bias---the ``integral constraint''---becomes.  (See \citet{1993ApJ...417...19H} for further discussion of bias due to the mean error and other approaches to avoid it.) 

In the appendix, we compute the bias explicitly.  We further show that the ensemble average of the naive, biased estimator may be cast as a linear operation applied to the true correlation function:
\begin{equation}
 \langle \tilde C_0(\theta_p) \rangle = \sum_{q} M_{pq} C(\theta_q),
\end{equation}
or as a matrix equation,
\begin{equation}
 \langle \mathbf{\tilde C_0} \rangle = \mathbf{M C}.
\end{equation}
{Writing it this way is somewhat analogous to the MASTER technique \citep{2002ApJ...567....2H} for CMB power spectrum estimation on the partial sky.}

In the appendix we find that the matrix is
\begin{equation}
  M_{pq} = \delta_{pq} - 2 \frac{\sum_{ij} d_{ij}(\theta_p) \alpha_i \alpha_jD^{(1)}_{iq}}{\sum_{ij} d_{ij}(\theta_p) \alpha_i \alpha_j} + D_{q}^{(2)} \tag{\ref{eqn:matrix_cont}}
\end{equation}
where the auxiliary operations
\begin{equation}
  D_{iq}^{(1)} = \frac{\sum_{k} \alpha_k  d_{ik}(\theta_q) }{\sum_k \alpha_k}
\qquad \qquad D_{q}^{(2)} =  \frac{\sum_{kl} \alpha_l \alpha_k d_{kl}(\theta_q) }{\left(\sum_k \alpha_k\right)^2} \tag{\ref{eqn:D1}, \ref{eqn:D2}}
\end{equation}
are functions of the pixel weights.  This matrix is composed of three terms.  The first term is the identity matrix and the following two terms are responsible for the bias.  The matrix costs ${\cal O}(N^2)$ operations to compute, the same as the naive correlation function estimator.

\subsection{Monte Carlo simulation}

To test our expression for the bias terms, we performed Monte Carlo simulations of continuous, diffuse fields; later we will include shot noise.  The survey size,  roughly $7' \times 8'$, mimics an actual observation with Chandra.  For the correlation function $C(\theta)$ in the simulation, we use a Gaussian function with correlation length (i.e. standard deviation) of $3.9'$, significant compared to the size of the field. For weights we use the inverse of the exposure for a real set of observations.  These downweight the edges of the observations compared to the center (and correspond to inverse-variance pixel weights in the Poisson-noise-dominated limit.)

\begin{figure}
\begin{center}
\includegraphics[width=0.6\columnwidth]{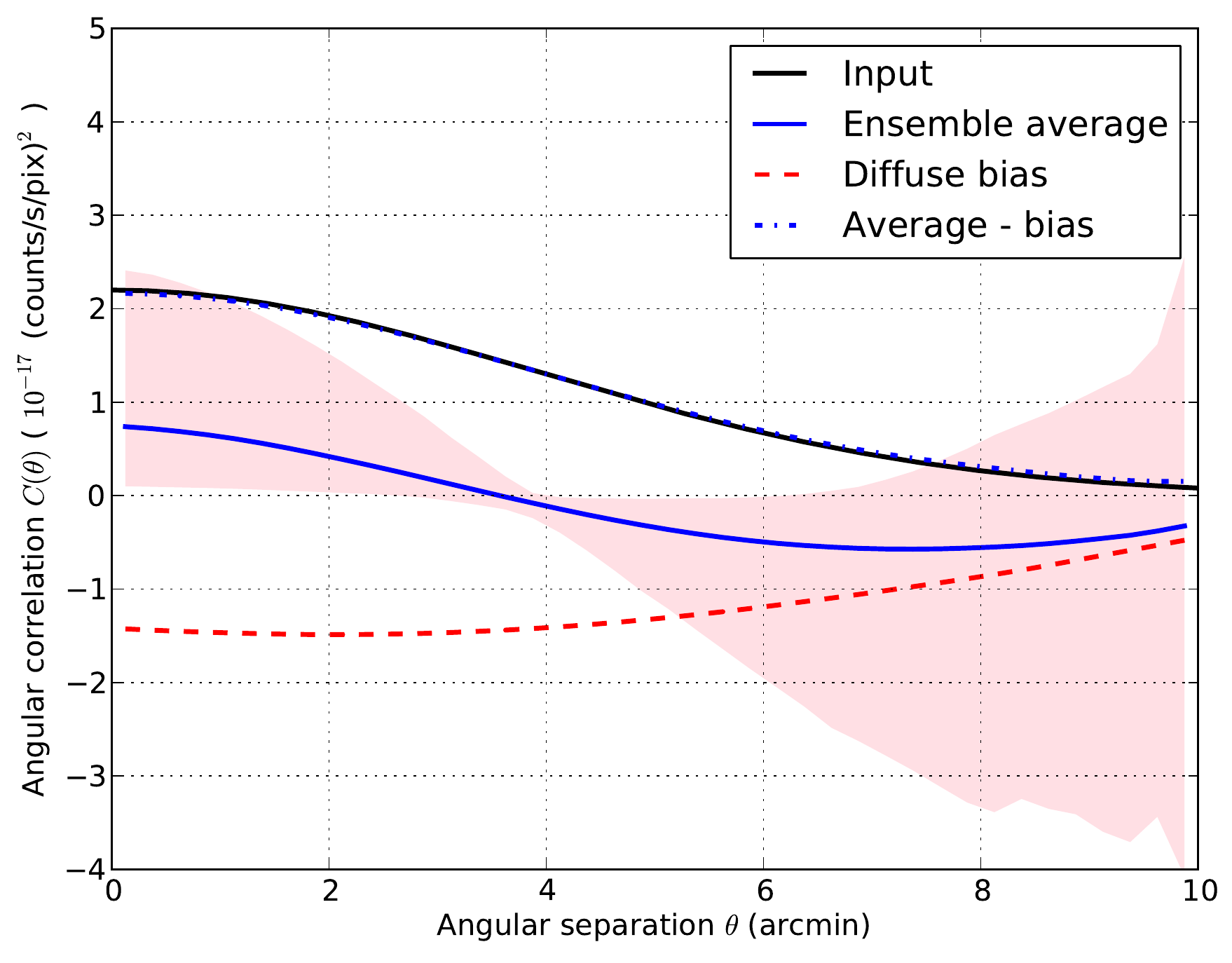}
\end{center}
\caption{The input correlation function (black) was used to create a set of $N_{\rm MC} = 1000$ Monte Carlo realizations of a simulated map (without shot noise).  At each angular separation, 95 percent of naive estimates $\tilde C_0(\theta)$ for the correlation function fall within the pink region.  The average of the Monte Carlo ensemble of naive estimates is solid blue, and has fluctuations reduced by a factor $\sqrt{N_{\rm MC}} \sim 30$. The sum of the bias terms computed from the input $C(\theta)$ is shown as the dashed red line.  The ensemble average minus the bias terms is shown with the dash-dot blue line, and closely matches the input.}
\label{fig:Wtheta_MC}
\end{figure}
\begin{figure}
\begin{center} 
\includegraphics[width=0.49\columnwidth]{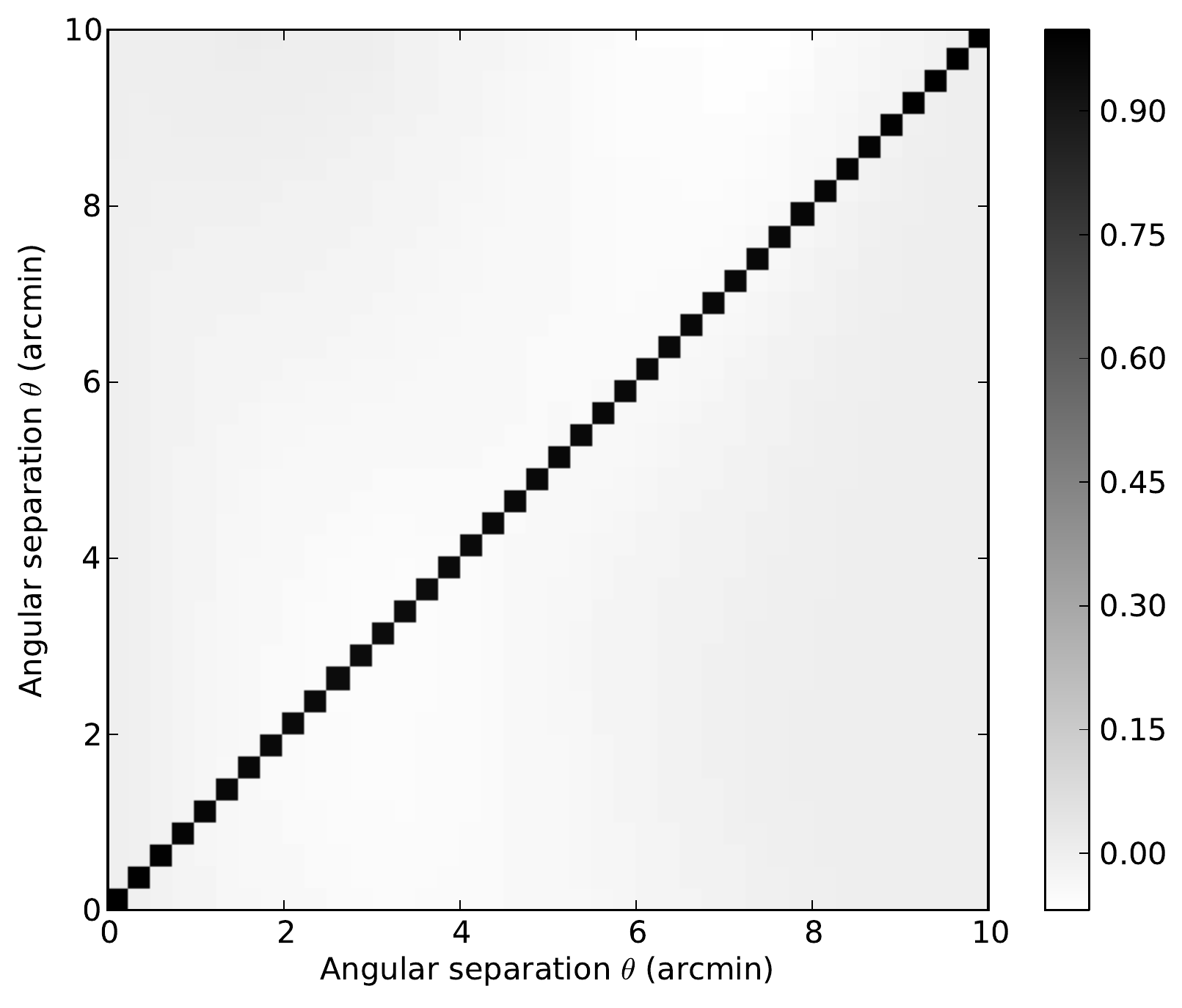}
\includegraphics[width=0.49\columnwidth]{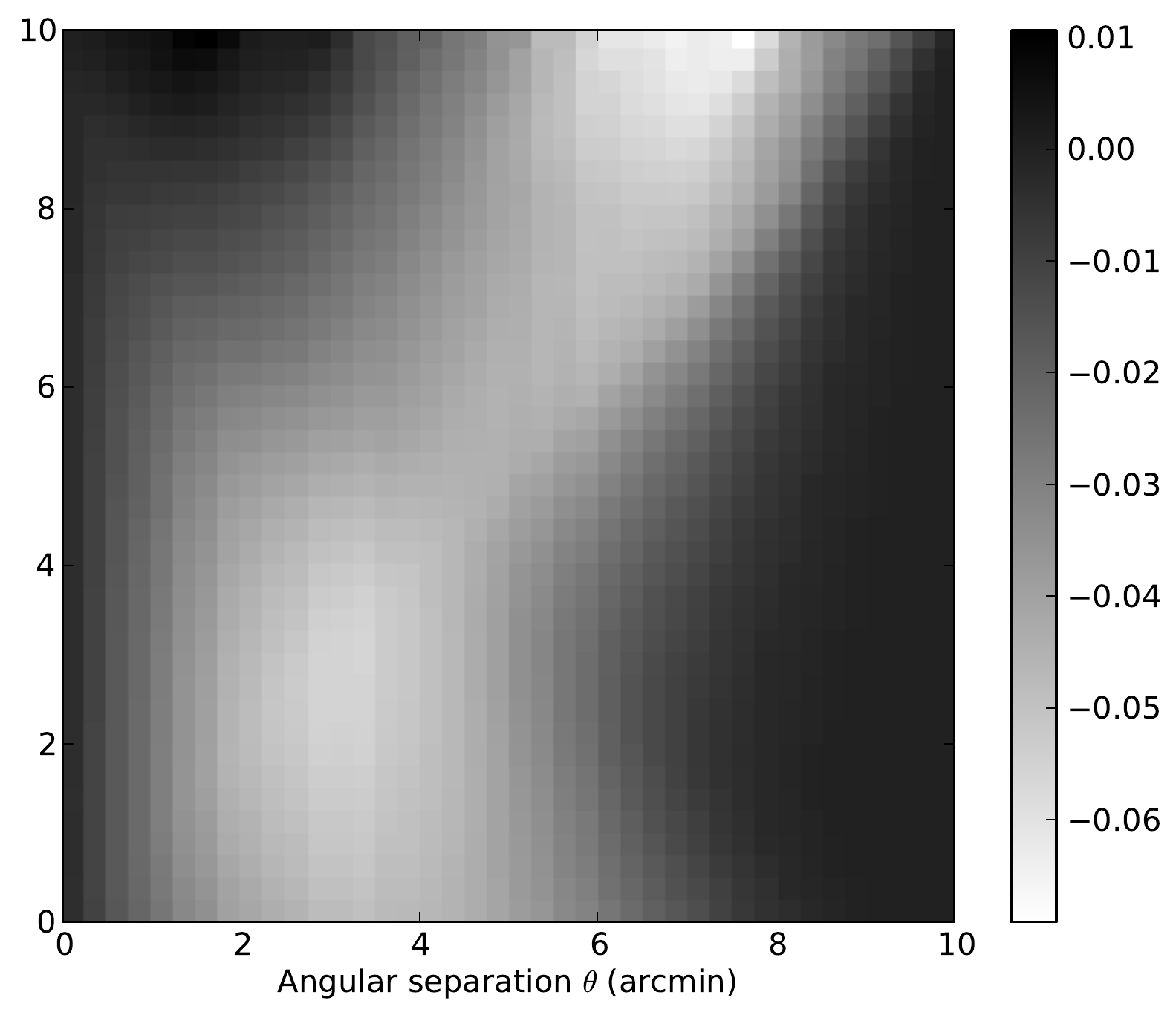}
\end{center} 
\caption{Left: the matrix $\mathbf M$ which relates the true correlation function to the ensemble average of the naive estimate.  The columns represent the input scale and the rows the output scale.  The matrix is dimensionless.  Right:  Without the identity matrix, we have the biasing terms only.}
\label{fig:biasmatrix}
\end{figure}

Figure \ref{fig:Wtheta_MC} shows the input correlation, and the ensemble average (and dispersion) of the naive estimates, which are biased.  Compared to the input correlation, the ensemble average is offset and the shape differs.  The bias terms capture this difference, but the bias terms depend on the input correlation function, and so when working with data are not directly available.  We address this shortcoming in the next section. The matrix $\mathbf{M}$ for our example is depicted in Figure \ref{fig:biasmatrix}.

 In the simulations shown, we generated the diffuse signal $s$ as a Gaussian random field, but obtain the same results with a log-normal random field (constructed with the recipe from \citet{2012ApJ...750...28C} to keep the same mean and correlation function).  The ensemble average and bias terms are the same in the Gaussian and non-Gaussian cases, however the non-Gaussianities substantially increase the dispersion of the naive estimates.

\subsection{Correcting the naive estimator}  \label{sec:continuous_svd}
 Once we have $\mathbf{M}$, we can define a \textit{reconstructed correlation function} $\tilde C(\theta_q)$ as the solution to the linear equation 
\begin{equation}
 \tilde C_0(\theta_p) = \sum_{q} M_{pq} \tilde C(\theta_q),
\label{eq:reconstructed}
\end{equation}
where the left-hand-side is the naive estimate we already obtained and the right-hand-side contains our reconstruction.

Unfortunately this equation does not have a unique solution.  Explicit computation in the appendix shows that $\mathbf{M}$ maps any constant offset to zero.  Thus constant offsets to the correlation function are in the null space of the matrix.  In particular this implies that $\mathbf{M}$ is not invertible, ruling out a straightforward solution to the linear equation.  However,
we can recover the true $C(\theta)$ in the ensemble average up to an unknown constant function.

Since we know this matrix has a non-empty null space, we analyze it by singular value decomposition, factoring it as
\begin{equation}
  \mathbf{M = U s V}^T
\end{equation}
where $\mathbf{U}$ and  $\mathbf{V}$ are orthogonal and $\mathbf{s}$ is diagonal and contains the singular values.  The matrix has one singular value near zero, and the column of $\mathbf{V}$ that corresponds to the singular mode contains the constant function we identified previously as being in the null space.  

The upshot of this discussion is that although $\mathbf{M}$ does not have an inverse, we can construct a pseudo-inverse 
\begin{equation}
  \mathbf{M^+ = V s^+ U}^T
\end{equation}
where $\mathbf{s^+}$ is a diagonal matrix constructed from the reciprocal of the diagonal of $\mathbf{s}$ except at the singular value where it is set to zero.  Then the reconstructed correlation function 
\begin{equation}
\tilde C(\theta_p) = \sum_q M^+_{pq} \tilde C_0(\theta_q)
\end{equation}
solves equation~(\ref{eq:reconstructed}).  This solution is not unique, however, since adding any constant function also yields a solution.  This procedure chooses the solution which minimizes the squared norm of the reconstructed correlation function \citep[e.g.][]{1992nrca.book.....P}
\begin{equation}
\sum_p | \tilde C(\theta_p) |^2.
\end{equation}
\begin{figure} 
\begin{center}
\includegraphics[width=0.6\columnwidth]{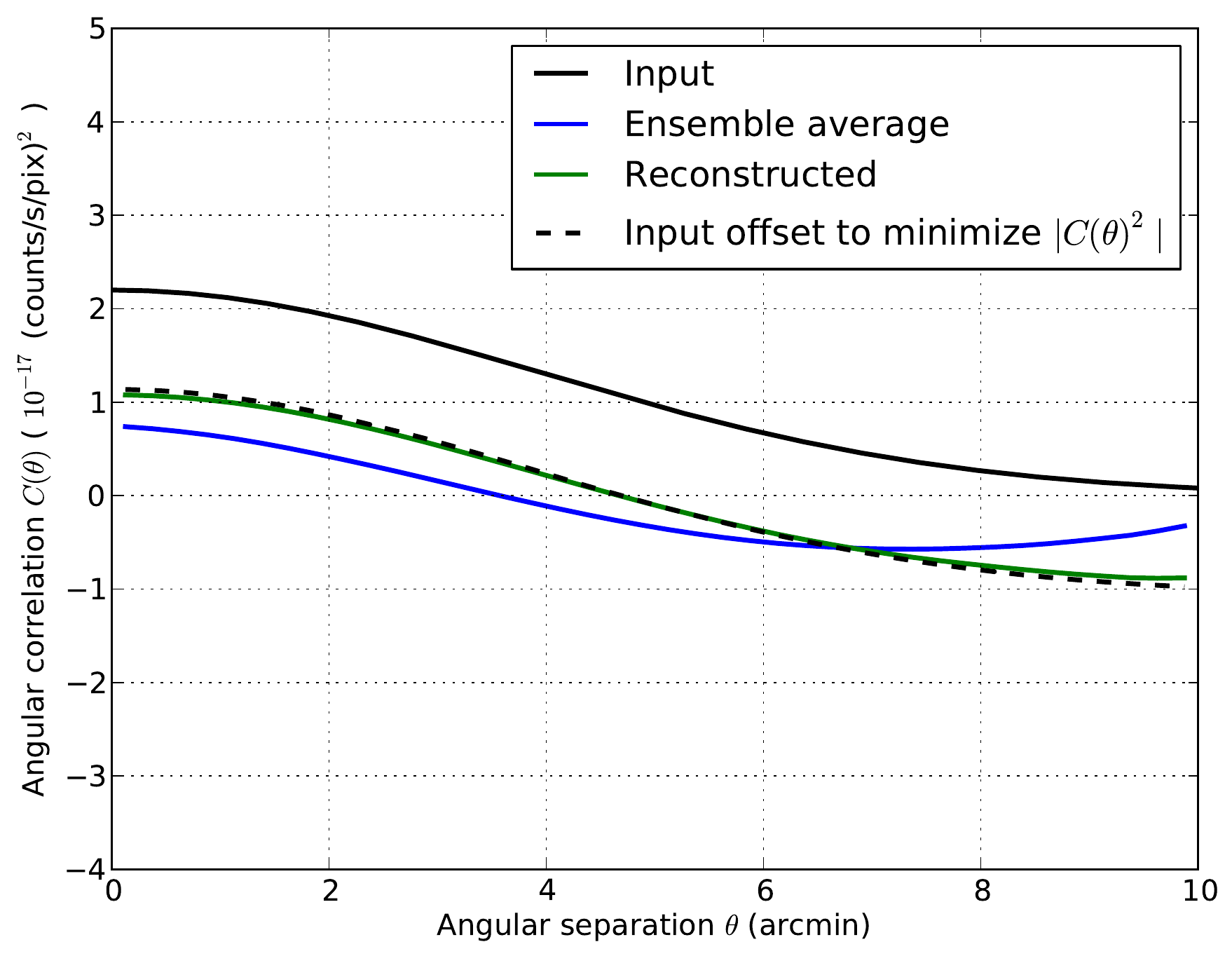}
\end{center}
\caption{The ensemble of 1000 realizations made with the input correlation shown in black yields the average naive correlation function shown in blue. 
Multiplying the ensemble average by  $\mathbf M^+$, the pseudo-inverse of the biasing matrix, gives the reconstructed correlation function (in green), which has the same shape as the input spectrum, but has lost the information about the constant offset.  It resembles the input spectrum after the input is offset to minimize the square norm.}
 \label{fig:Wtheta_MC_reconstruction}
\end{figure}
Therefore, in the ensemble average, we can reconstruct the correlation matrix up to a constant offset factor, as shown in Fig.~\ref{fig:Wtheta_MC_reconstruction} for our Monte Carlo simulation.  This shows how the incorrect shape of the ensemble average has been repaired in the reconstruction, except for residual fluctuations in the ensemble average.

Thus we have 
\begin{equation}
 \langle \tilde C(\theta_p) \rangle = C(\theta_p) + \mbox{const.}
\end{equation}
where the constant is unknown.  Our estimator is therefore biased.  Note however, that the shape is not biased, as we can see from a comparison of the reconstructed correlation function at two separations:
\begin{equation}
 \langle \tilde C(\theta_p) - \tilde C(\theta_q) \rangle = C(\theta_p) + \mbox{const.} - C(\theta_q) - \mbox{const.} = C(\theta_p) - C(\theta_q)
\end{equation}
for any scales $\theta_p$ and $\theta_q$ accessible by the survey.  Thus we can say that the shape information is preserved in an unbiased way.
If we further have theoretical expectations or other constraints, these can help fix the offset for the correlation function.

\section{Poisson shot noise}\label{sec:poisson}

If the observations have significant shot noise from measuring discrete photons or objects, additional bias terms appear.  We use a Poisson model \citep[][\S 33]{1980lssu.book.....P} for our computations.  Let $N_i$ be the count of events in pixel or cell $i$.  This quantity is Poisson-distributed with a mean parameter $\lambda_i$ that is proportional to our diffuse signal.  In our X-ray example, $\lambda_i = s_i t_i A$, where $s_i$ is our diffuse signal from before, representing a photon rate per area, time $t_i$ is the duration of the pixel's exposure, and $A$ is the pixel's collecting area.\footnote{These may differ for other applications. For the example of galaxy counts, the galaxy number density plays the role of the signal and the cell volume plays the role of the exposure-weighted area.}  Note $\lambda_i$ is a mean number of counts, and so is dimensionless.  The Chandra observations we have studied have a large fraction of counts ($\sim 85$ percent) that are spurious events unrelated to the cosmic signal.  We first derive the bias and corrections for the naive estimator neglecting these spurious counts, and then including them.

\subsection{No spurious contamination}\label{sec:nobackground}

If all the counts are genuinely related to the cosmic signal, the observed rate ($R$) of signal events is 
\begin{equation}
  R_i = N_i/t_i A
\end{equation}
which has the same units as $s_i$.  The ensemble average of $R_i$ is 
\begin{equation}
  \langle R_i \rangle =  \frac{\langle N_i \rangle}{t_i A} =  \frac{\langle s_i \rangle t_i A}{t_i A} = \mu.
\end{equation}
We can estimate the mean of our rate map
\begin{equation}
  \bar R = \frac{\sum_i \alpha_i R_i}{\sum_i \alpha_i}
\end{equation}
which is an unbiased estimate: $\langle \bar R \rangle = \mu$.  The fluctuation in the map's mean we call
\begin{equation}
  \delta \bar R = \bar R - \mu
\end{equation}
which has  $\langle \delta \bar R \rangle = 0$.  The covariance of the observed rate map is
\begin{eqnarray}
  {\rm Cov}(R_i,R_j) 
  &=& \frac{\mu}{t_iA} \delta_{ij} + C(\theta_{ij}). \label{eq:poissonratecov}
\end{eqnarray}
This has an additional shot noise component compared to the covariance of the diffuse signal.  The shot noise term can be avoided if the sums over pixel pairs exclude common pixels, at the cost of slightly more complicated pixel accounting.  Here we include it in our computations for completeness.

Note that since $C(\theta)$ is a property of the diffuse field's probability distribution, in the discrete case it is not subject to any particular new constraints compared to the continuous case.  The total number of counts (or objects) summed over all pixels is a random variable, and is \textit{not} fixed \citep[][cf. \S 31, 33 vs. \S 32]{1980lssu.book.....P}, and there is no specific constraint on the integral of $C(\theta)$.

The field $s$, representing a rate of counts or objects, must be non-negative, which implies that its statistics are non-Gaussian.  For the derivation of the estimator biases, this matters little because, as before, the higher-order moments do not appear in our argument.  On the other hand, it may matter more when constructing simulations.  A Gaussian random field can be a suitable approximation for $s$, but only if the particular realizations do not contain negative pixels, which would lead to negative (and thus ill-defined) expected counts.  Otherwise, a log-normal random field, which is positive-definite and which we employ below, provides another useful candidate.

As before we make a naive estimate of the correlation function
\begin{equation}
\tilde C_0^R(\theta) =  \frac{ \sum_{ij} d_{ij}(\theta) \alpha_i \alpha_j (R_i-\bar R)(R_j - \bar R)}{\sum_{ij} d_{ij}(\theta) \alpha_i \alpha_j}  \label{eqn:poisson_naive}.
\end{equation}

In the appendix, we show that the ensemble average of the naive estimator for the discrete field can be written as a linear function of both the true mean and the true correlation function.
\begin{equation}
\langle \tilde C_0^R(\theta_p) \rangle  = v^R_p \mu + \sum_q M_{pq} C(\theta_q) \tag{\ref{eqn:bias_poisson}}
\end{equation}
where 
\begin{gather}
v^R_p = \frac{ \sum_{ij} d_{ij}(\theta_p) \alpha_i \alpha_j [(1/t_iA)\delta_{ij} - 2 E^{(1)}_i + E^{(2)}]}{\sum_{ij} d_{ij}(\theta_p) \alpha_i \alpha_j} \nonumber\\
E^{(1)}_i = \frac{\alpha_i/t_iA}{\sum_k \alpha_k} 
\qquad \qquad 
E^{(2)} =  \frac{\sum_k \alpha_k^2/t_kA}{(\sum_k \alpha_k)^2} 
\tag{\ref{eqn:E1}, \ref{eqn:E2}, \ref{eqn:vR}}
\end{gather}
and $\mathbf{M}$ is the same matrix as before.

We can express this relationship in matrix form as 
\begin{equation}
\left( 
  \begin{array}{c}
    \langle \bar R \rangle \\
    \langle \mathbf{ \tilde C}_0^R \rangle 
  \end{array}
\right) 
= \left(
\begin{array}{cc}
  1 & (0 \dots 0) \\
  \mathbf{v}^R &   \mathbf{M} 
\end{array}
\right)
\left( 
  \begin{array}{c}
    \mu \\
    \mathbf{C}
  \end{array}
\right) 
\end{equation}
where we used that $\bar R$ is an unbiased estimator for $\mu$.

Like  $\mathbf M$ before, this larger square matrix is amenable to the construction of a pseudo-inverse by singular value decomposition.  Analogous to equation (\ref{eq:reconstructed}), we can solve the linear equation 
\begin{equation}
\left( 
  \begin{array}{c}
    \bar R  \\
    \mathbf{ \tilde C}_0^R  
  \end{array}
\right) 
= \left(
\begin{array}{cc}
  1 & (0 \dots 0) \\
  \mathbf{v}^R &   \mathbf{M} 
\end{array}
\right)
\left( 
  \begin{array}{c}
    \tilde \mu \\
    \mathbf{ \tilde C}
  \end{array}
\right) 
\end{equation}
to reconstruct estimates (on the right-hand side) for the mean (this estimate is unbiased because it just takes the already unbiased $\bar R$ directly) and correlation function, with the same limitation as before: a constant function added to the correlation function is unconstrained.  As before, the shape of the reconstructed correlation function in the ensemble average matches the true correlation function.

\subsection{With spurious contamination}\label{sec:background}
In the presence of an uncorrelated, but spatially varying, set of spurious counts, the analysis changes slightly, with the spurious counts contributing additional shot noise terms.  In the case of Chandra data, these spurious counts are well-characterized in the sense that their mean rate is well-understood.  However, counts cannot be classified as signal or spurious on an individual basis.

Now our counts include events from both the signal and the spurious set: $N_i = N_i^s + N_i^{sp}$.  Then the ensemble average photon count is $\langle N_i \rangle = \mu t_i A + \lambda^{sp}_i$, where $\lambda_i^{sp}$ is the known spurious mean count for each pixel.  We redefine the signal rate map as
\begin{equation}
  R_i = \frac{N_i - \lambda^{sp}_i}{t_iA} \label{eqn:rate_spur}
\end{equation}
so that $\langle R_i \rangle = \mu$.  Defining the map mean as before yields $\langle \bar R \rangle = \mu$ and the fluctuation from the mean has average $\langle\delta \bar R\rangle = 0$.
From here the analysis proceeds much as before.  Noting that 

\begin{equation}
  \langle (N_i - \lambda^{sp}_i)(N_j - \lambda^{sp}_j) \rangle = \langle N_i^s N_j^s \rangle + {\rm Cov}(N_i^{sp},N_j^{sp})
\end{equation}
we can show that
\begin{equation}
  {\rm Cov}(R_i,R_j) = \left( \frac{\mu}{t_iA} + \frac{\lambda^{sp}_i}{t_i^2A^2} \right)\delta_{ij} + C(\theta_{ij}). \label{eqn:cov_spur}
\end{equation}
 which includes an additional shot noise term compared to the similar eqn.~(\ref{eq:poissonratecov}).

This allows the ensemble average of the naive estimate to be written as the sum of the spurious-event-free naive estimate and additional shot-noise terms which depend on the known mean spurious rate, $\lambda_i^{sp}$.  In the appendix we show that this is:
\begin{gather}
  \langle \tilde C_0^{R,sp}(\theta) \rangle =  \langle \tilde C_0^{R}(\theta) \rangle +
   \frac{ \sum_{ij} d_{ij}(\theta) \alpha_i \alpha_j (\lambda_i^{sp}/t_i^2A^2)\delta_{ij} - 2\left(\alpha_i\lambda_i^{sp} /(t_i^2A^2 \sum_k \alpha_k) \right)}{ \sum_{ij} d_{ij}(\theta) \alpha_i \alpha_j} + \frac{\sum_k \alpha_k^2 \lambda_k^{sp}/t_k^2}{A^2\left( \sum_k\alpha_k \right)^2}
\tag{\ref{eqn:ensemble_spur}}
\end{gather}
Subtracting away these spurious terms, we can proceed to reconstruct the correlation function as described at the end of section \ref{sec:nobackground}.

\subsection{Poisson Monte Carlo simulation}

\begin{figure}
\begin{center}
\includegraphics[width=0.6\columnwidth]{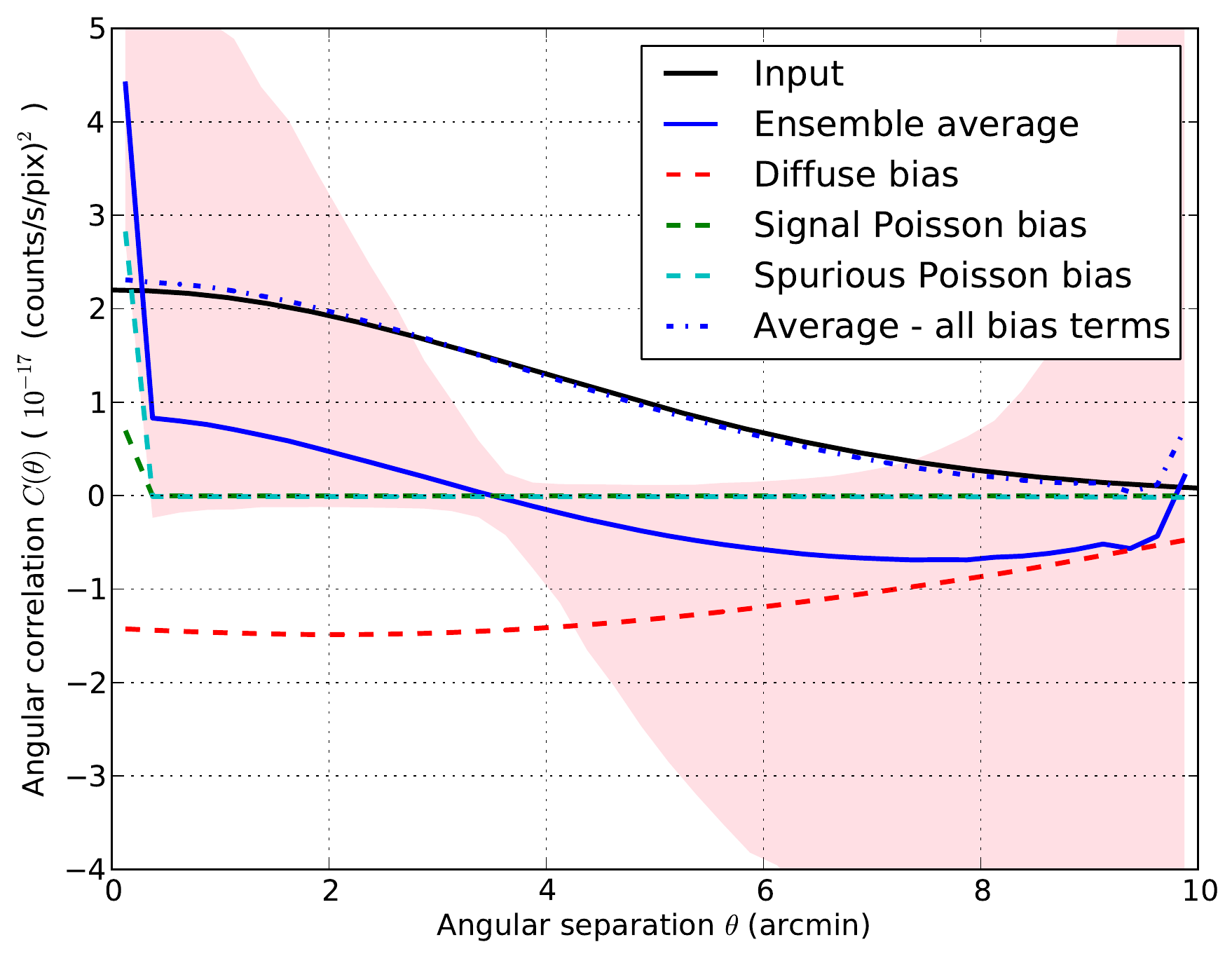}
\end{center}
\caption{Similar to figure \ref{fig:Wtheta_MC}, except including shot noise from signal photons and background events, based on 5000 log-normal random fields.  The Poisson bias terms (dashed green and cyan) are very small except in the first bin, which contains common pixel pairs. Accounting for all bias terms, the average closely matches the input, including at the first bin.}
\label{fig:Wtheta_obs_MC}
\end{figure}

For a set of 5000 Monte Carlo realizations that include shot noise, we show in Fig.~\ref{fig:Wtheta_obs_MC} the ensemble average and dispersion for the naive estimate, and also the analytic computation of the bias terms.
The mean rate of photons, $\mu = 4.3 \times 10^{-9}$ counts/s/pixel, was chosen based on a real Chandra observation, and is low enough that a Gaussian random field with this correlation function will have negative pixels.  For this reason we used a log-normal random field in this case, which accounts for much of the increase in the dispersion compared to  Fig.~\ref{fig:Wtheta_MC}.
The shot-noise bias terms are large in the first bin of the correlation function, which contains the same-pixel pairs.  Elsewhere, they are small because in this application, we have enough photons to make the shot noise contribution to $\delta \bar R$ sub-dominant.  The bias terms we computed account for the shot noise well.  The dispersion due to shot noise is extreme at $>9'$ separations for two reasons: only the periphery of the map provides these separations, so there are few pixel pairs, and the effective exposure for pixels at the edge of the map is less, so there are many fewer photons than at the center of the field.

\begin{figure}
\begin{center}
\includegraphics[width=0.6\columnwidth]{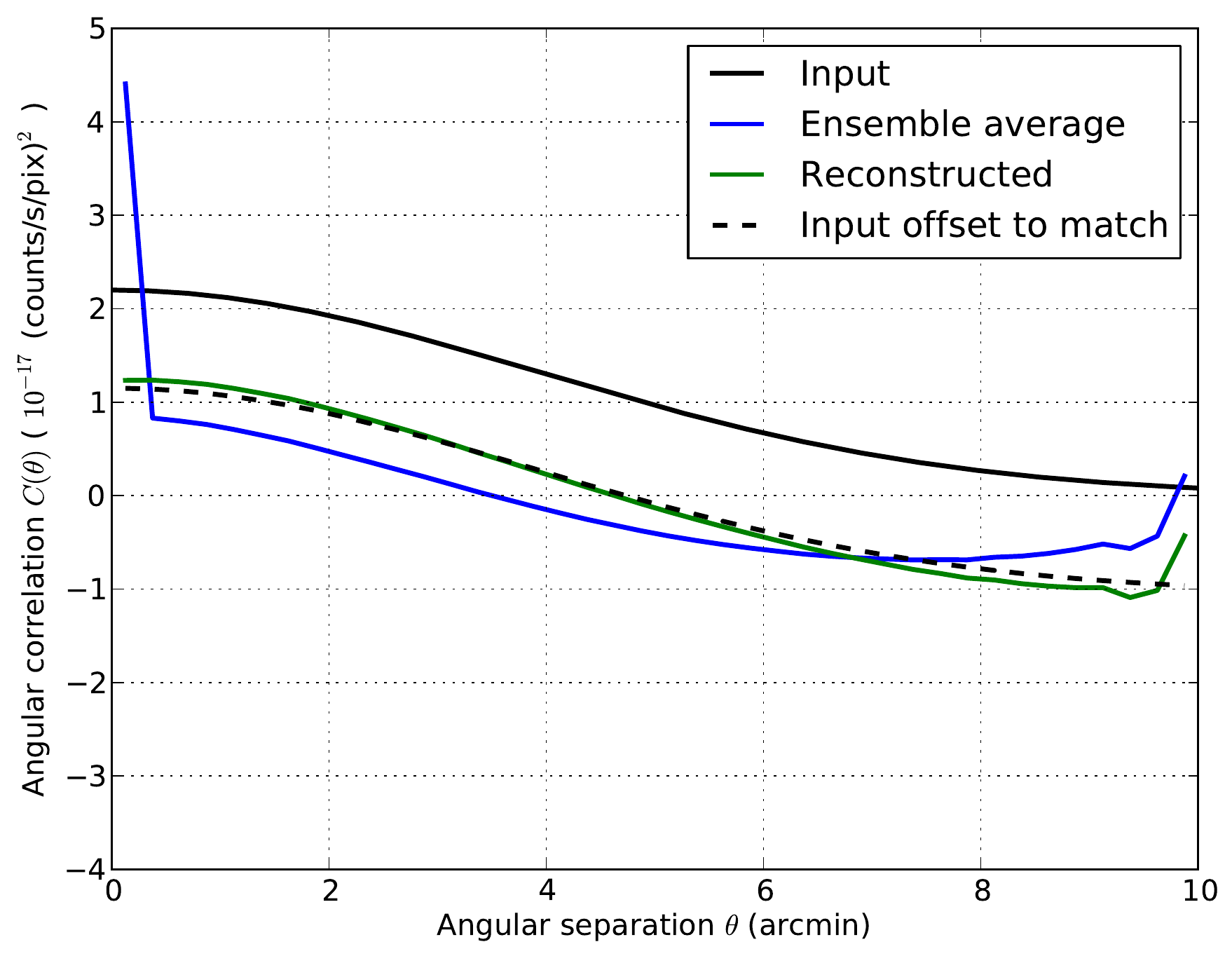}
\end{center}
\caption{Similar to figure \ref{fig:Wtheta_MC_reconstruction}, reconstructing the correlation function, except including shot noise from signal photons and background events.}
\label{fig:Wtheta_MC_poisson_reconstruction}
\end{figure}

In Fig.~\ref{fig:Wtheta_MC_poisson_reconstruction}, we demonstrate that the reconstruction of the correlation function by the singular value decomposition method works well to correct the shape distortion in the ensemble average.

\section{Conclusions} \label{sec:conclusions}

We have developed an estimator for the correlation function which allows the shape, but not the overall offset, of the correlation function to be estimated properly in the ensemble average.  If there are significant signal correlations on the largest scales that the survey region can probe, as with X-ray observations and some other astronomical data sets, the large sample variance will limit the utility of the correlation function shape measurement.  However, when $\tilde C(\theta)$'s from multiple fields are averaged, we beat down the noise on the shape, while the average of unknown offsets simply yields a new unknown offset.  Put another way, averaging improves our knowledge of the shape but does not worsen our lack of knowledge about the offset.

The estimators written here, although motivated by observations of the diffuse X-ray background, easily generalize to galaxy counts-in-cells (setting $\lambda_i = {\cal N}_i\Delta\Omega$ in section \ref{sec:poisson}).  The estimator can be trivially adapted for cross-correlations between fields, or extended from angular correlations in two dimensions  to linear or time-series correlations in one dimension or spatial correlations in three dimensions.

These estimators may be usefully applied to any situation with correlations on the scale of the observed region.  One example is the CMB, which in the $\Lambda$CDM model has significant correlations even between points on the sky separated by $180^\circ$.  However, estimates from the COBE and WMAP data \citep{1996ApJ...464L..25H,2003ApJS..148..175S,2007PhRvD..75b3507C,2009MNRAS.399..295C} show surprisingly little correlations at scales larger than $60^\circ$.  These authors have used the biased, naive estimator (equation~\ref{eq:Cestimator}),  but our preliminary tests on WMAP maps and the $\Lambda$CDM CMB correlation function indicate that the bias terms we have computed here are too small to account for this difference. 
 
We have computed the variance of our estimates in Monte Carlo simulations, but not analytically, nor have we tried to find optimal weights to minimize the variance.  When sample variance dominates the covariance for the correlation function, it is unlikely that the optimal weighting can be done on a pixel-by-pixel basis, and instead pixel pairs will need to be jointly weighted by the inverse covariance for that pair, accounting for the signal covariance and the signal and spurious shot noise. Compared to the real-space estimators we examine here, \citet{2004MNRAS.348..885E} and \citet{2010MNRAS.407.2530E} argue that a correlation function estimate built from a maximum likelihood estimate of the harmonic space power spectrum will have lower variance, because it effectively gives pixel pairs closer-to-optimal weights in this way.  This task we leave for future work.

\section*{Acknowledgments}
We thank Enzo Branchini and the anonymous referee for useful comments on earlier versions of this work.  We thank Gabriela Degwitz for help in the preparation of this manuscript.
This work was supported by NASA through the Smithsonian Astrophysical Observatory
(SAO), award G0112177X, and NASA award NNX11AF80G.
KMH also receives support from NASA-JPL subcontract 1363745.

\bibliographystyle{apj}
\bibliography{notes.bib}

\appendix

\section{Bias terms: continuous case} \label{sec:bias_cont}

In this appendix we compute the bias terms for the continuous signal.  
Rewriting $ \tilde\mu = \mu + \delta\tilde\mu$, the ensemble average of the numerator of the naive estimator (\ref{eq:Cestimator}) is
\begin{equation}
\sum_{ij} d_{ij}(\theta_p) \alpha_i \alpha_j \left[ C(\theta_{ij}) - \langle s_i\delta\tilde\mu \rangle  - \langle s_j\delta\tilde\mu \rangle + \langle \delta\tilde\mu^2 \rangle \right]
\end{equation}
where we have used $\langle \delta\tilde\mu \rangle = 0$.   Further we can use the sum's symmetry between $i$ and $j$ to show that it equals
\begin{equation}
\sum_{ij} d_{ij}(\theta_{p}) \alpha_i \alpha_j \left[ C(\theta_{ij}) - 2\langle s_i\delta\tilde\mu \rangle + \langle \delta\tilde\mu^2 \rangle \right]\label{eqn:numerator_ensemble_Average}.
\end{equation}
If we had used the true mean, only the $C(\theta_{ij})$ term would be present, and we could pull it out of the sum as $C(\theta_p)$. The sum over weights would cancel the denominator, and we would indeed find that $\langle \tilde C_0(\theta_p) \rangle = C(\theta_p)$.  This is not the case here because of the middle and last terms in the brackets, which are responsible for the bias.

We can compute both bias terms from the field's correlation function.  We call the first bias term $B^{(1)}_i$ because it is first order in the mean estimation error $\delta\tilde\mu$, and compute it as
\begin{eqnarray} \nonumber
  B^{(1)}_i = \langle  s_i\delta\tilde\mu \rangle &=&  \langle s_i(\tilde\mu - \mu) \rangle \\ \nonumber
  &=& \frac{\sum_k \alpha_k\langle s_i s_k\rangle}{\sum_k \alpha_k} - \mu^2 \\ \nonumber
  &=& \frac{\sum_k \alpha_k C(\theta_{ik})}{\sum_k \alpha_k}.
\end{eqnarray}
The second bias term, $B^{(2)}$, which is second order in the mean's error, has no dependence on the pixel index.
\begin{eqnarray}\nonumber
  B^{(2)} = \langle \delta\tilde\mu^2  \rangle&=&  \langle (\tilde\mu - \mu)^2 \rangle \\\nonumber
  & = & \left\langle \frac{\sum_k \alpha_ks_k}{\sum_k \alpha_k} \frac{\sum_l \alpha_ls_l}{\sum_l \alpha_l}   \right\rangle  - \mu^2 \\\nonumber
  &=& \frac{\sum_{kl} \alpha_l \alpha_k \langle s_k s_l \rangle}{\left(\sum_k \alpha_k\right)^2}  - \mu^2\\\nonumber
  & = &\frac{\sum_{kl} \alpha_l \alpha_k C(\theta_{kl}) }{\left(\sum_k \alpha_k\right)^2}
\end{eqnarray}
Because $B^{(2)}$ does not depend on the pixel index, this term too can slip outside the sum over pixel pairs in eqn.~(\ref{eqn:numerator_ensemble_Average}).  Therefore, finally, we have
\begin{equation}
  \langle \tilde C_0(\theta_p) \rangle = C(\theta_p) - 2 \frac{\sum_{ij} d_{ij}(\theta_p) \alpha_i \alpha_jB^{(1)}_i}{\sum_{ij} d_{ij}(\theta_p) \alpha_i \alpha_j} + B^{(2)} \label{eq:naivebias}
\end{equation}
which states the bias in our estimate explicitly.  Each bias term costs ${\cal O}(N^2)$ operations to compute, the same as the correlation function.

Note that our naive estimator has a peculiar reaction to correlation functions such as $C(\theta) = c$ for all separations sampled by our survey.\footnote{On scales larger than the survey, this correlation function could vary without changing the discussion.}  In this case $\langle \tilde C_0(\theta) \rangle = 0 $, which we show by examining the bias terms.  If $C(\theta) = c$, then the constant can be set outside the sums, which cancel the denominators.  Therefore bias factors $B_i^{(1)} = c$ and $B^{(2)} = c$, and the middle term of eqn.~(\ref{eq:naivebias}) is $-2c$.  Therefore  $\langle \tilde C_0(\theta) \rangle = c - 2c + c = 0$.  Thus, if the naive estimator is viewed as a linear operator on the input correlation function, constant functions are in the null space of the operator, since any constant maps to zero.  Moreover, the naive estimator loses the information about any constant baseline in the correlation function, although the information about the shape is preserved.

The bias terms depend on $C(\theta)$ only on scales accessible by the survey region, and not on any larger scales.  This permits an (imperfect) reconstruction of the correlation function.  To proceed, we can rewrite eqn.~(\ref{eq:naivebias}) as a matrix multiplication:
\begin{equation}
 \langle \tilde C_0(\theta_p) \rangle = \sum_{q} M_{pq} C(\theta_q)
\end{equation}
where the sum is over the angular bins.
Then we set about finding the matrix $\mathbf{M}$.

To write down $\mathbf{M}$, we make use of the relationship
\begin{equation}
  C(\theta_{ik}) = \sum_q d_{ik}(\theta_q) C(\theta_q).
\end{equation}
Note that this sum is over angular bin, not pixel.  We rewrite the bias terms more explicitly as linear operations on the vector $C(\theta_q)$.  The first bias term is
\begin{equation}
  B_i^{(1)} =  \frac{\sum_{kq} \alpha_k  d_{ik}(\theta_q) C(\theta_q)}{\sum_k \alpha_k} = \sum_q D_{iq}^{(1)} C(\theta_q),
\end{equation}
where we define
\begin{equation}
  D_{iq}^{(1)} = \frac{\sum_{k} \alpha_k  d_{ik}(\theta_q) }{\sum_k \alpha_k}.\label{eqn:D1}
\end{equation}
Note that the first index refers to pixel and the second to bin.  The second bias term is
\begin{equation}
  B^{(2)} = \frac{\sum_{kl} \alpha_l \alpha_k d_{kl}(\theta_q) C(\theta_q) }{\left(\sum_k \alpha_k\right)^2} =  \sum_q D_{q}^{(2)} C(\theta_q),
\end{equation}
where we define
\begin{equation}
 D_{q}^{(2)} =  \frac{\sum_{kl} \alpha_l \alpha_k d_{kl}(\theta_q) }{\left(\sum_k \alpha_k\right)^2}. \label{eqn:D2}
\end{equation}
Since $C(\theta_p) = \sum_q \delta_{pq} C(\theta_q)$, we finally have
\begin{equation}
  M_{pq} = \delta_{pq} - 2 \frac{\sum_{ij} d_{ij}(\theta_p) \alpha_i \alpha_jD^{(1)}_{iq}}{\sum_{ij} d_{ij}(\theta_p) \alpha_i \alpha_j} + D_{q}^{(2)} \label{eqn:matrix_cont}
\end{equation}

To sum up, in this appendix we have: (1) computed the bias of the naive correlation function estimator; (2) shown that the ensemble average of the naive estimate is a linear operation acting upon the true correlation; (3) computed that linear operator in terms of the pixel weights; and (4) shown that constant offsets are in the null space of that operator.  The method to estimate the shape of the correlation function in section \ref{sec:continuous_svd} depends on these results.

\section{Bias terms: discrete case} \label{sec:bias_discrete}

\subsection{No spurious contamination}
To compute the bias for the discrete case, we write the numerator of the naive estimator (\ref{eqn:poisson_naive}) in terms of the fluctuation of the mean $\delta \bar R$ and take the ensemble average:
\begin{eqnarray}
&&  \left\langle\sum_{ij} d_{ij}(\theta) \alpha_i \alpha_j (R_i-\mu - \delta\bar R)(R_j - \mu - \delta \bar R) \right\rangle \nonumber \\
&=&  \left\langle \sum_{ij} d_{ij}(\theta) \alpha_i \alpha_j \left[ (R_i-\mu)(R_j-\mu) - 2(R_i - \mu) \delta\bar R + (\delta\bar R)^2 \right] \right\rangle \nonumber \\
&=& \sum_{ij} d_{ij}(\theta) \alpha_i \alpha_j \left[(\mu/t_iA)\delta_{ij} + C(\theta_{ij}) -2 \langle R_i \delta\bar R \rangle + \langle (\delta\bar R)^2 \rangle   \right]
\end{eqnarray}
Now we examine the last two terms, which are analogous to the bias terms for the diffuse signal.  First,
\begin{eqnarray}
B^{R(1)}_i = \langle R_i \delta\bar R \rangle &=& \frac{\sum_k \alpha_k \langle R_i R_k \rangle}{\sum_k \alpha_k} - \mu^2 \nonumber \\
&=& \frac{\sum_k \alpha_k [(\mu/t_iA) \delta_{ik} + C(\theta_{ik})]}{\sum_k \alpha_k} \nonumber \\
&=& E^{(1)}_i \mu + B^{(1)}_i   \label{eq:BR1}
\end{eqnarray}
where we define 
\begin{equation}
E^{(1)}_i = \frac{\alpha_i/t_iA}{\sum_k \alpha_k}. \label{eqn:E1}
\end{equation}
This shows that for a signal of discrete photons, this bias term can be written as a sum of a new shot noise term and the old $B^{(1)}$ bias term from the diffuse case.

The final term is 
\begin{eqnarray}
  B^{R(2)} =  \langle (\delta\bar R)^2 \rangle &=&  \frac{\sum_{kl} \alpha_k \langle R_k R_l \rangle}{\sum_{kl} \alpha_k \alpha_l} - \mu^2  \nonumber \\
&=&   \frac{\sum_{kl} \alpha_k \alpha_l [(\mu/t_iA)\delta_{kl} + C(\theta_{kl})]}{(\sum_{k} \alpha_k)^2} \nonumber \\
&=&E^{(2)}\mu + B^{(2)} \label{eq:BR2}
\end{eqnarray}
where we define
\begin{equation}
  E^{(2)} =  \frac{\sum_k \alpha_k^2/t_kA}{(\sum_k \alpha_k)^2}. \label{eqn:E2}
\end{equation}
Again this bias term has a new, shot-noise component added to the old bias term from the diffuse signal.  These shot noise bias terms cannot be avoided by excluding $i=j$ from the naive estimator's pixel sums.

Each of the new shot noise terms is proportional to $\mu$.  We can gather those terms together and notice that the remaining terms are just those which appear on the right side of eqn.~(\ref{eq:naivebias}), so that:
\begin{equation}
\langle \tilde C_0^R(\theta_p) \rangle =  \left[\frac{ \sum_{ij} d_{ij}(\theta_p) \alpha_i \alpha_j [(1/t_iA)\delta_{ij} - 2 E^{(1)}_i + E^{(2)}]}{\sum_{ij} d_{ij}(\theta_p) \alpha_i \alpha_j} \right] \mu + \langle \tilde C_0(\theta_p) \rangle \label{eqn:vR}
\end{equation}
Therefore the ensemble average of the naive estimate for the discrete signal equals the ensemble average of the naive estimate for the diffuse signal plus an additional shot noise bias term which is proportional to the mean of the diffuse field.

Thus the ensemble average of the naive estimator for the discrete field can be written as a linear function of the true mean and correlation function.
\begin{equation}
\langle \tilde C_0^R(\theta_p) \rangle  = v^R_p \mu + \sum_q M_{pq} C(\theta_q). \label{eqn:bias_poisson}
\end{equation}
This formulation leads to the reconstruction method for the correlation function discussed in section \ref{sec:nobackground}.

\subsection{Including spurious contamination}

Starting from equations (\ref{eqn:rate_spur}) and (\ref{eqn:cov_spur}), we find that the two bias terms also have additional shot noise components due to the spurious signal.  Instead of eqn.~(\ref{eq:BR1}) we have
\begin{equation}
B_i^{R(1)} = \langle R_i \delta\bar R \rangle = E_i^{(1)}\mu + B_i^{(1)} + \frac{\alpha_i\lambda^{sp}_i}{t_i^2A^2\sum_k \alpha_k},
\end{equation}
 and instead of eqn.~(\ref{eq:BR2}) we have
\begin{equation}
B^{R(2)} = \langle (\delta\bar R)^2 \rangle = E^{(2)}\mu + B^{(2)}  + \frac{\sum_k \alpha_k^2 \lambda_k^{sp}/t_k^2}{A^2\left( \sum_k\alpha_k \right)^2}.
\end{equation}

Thus there are additional terms which can be subtracted away to yield the naive estimator in the contamination-free case.
\begin{gather}
  \langle \tilde C_0^{R,sp}(\theta) \rangle =  \langle \tilde C_0^{R}(\theta) \rangle + 
   \frac{ \sum_{ij} d_{ij}(\theta) \alpha_i \alpha_j (\lambda_i^{sp}/t_i^2A^2)\delta_{ij} - 2\left(\alpha_i\lambda_i^{sp} /(t_i^2A^2 \sum_k \alpha_k) \right)}{ \sum_{ij} d_{ij}(\theta) \alpha_i \alpha_j} + \frac{\sum_k \alpha_k^2 \lambda_k^{sp}/t_k^2}{A^2\left( \sum_k\alpha_k \right)^2}
\label{eqn:ensemble_spur}
\end{gather}

\end{document}